\documentclass[aps,twocolumn,showpacs]{revtex4}

\begin{document}

\title{Perfect Correlations between Noncommuting
Observables}
\author{Masanao Ozawa}
\affiliation{Graduate School of Information Sciences, T\^{o}hoku University,
Aoba-ku,  Sendai, 980-8579, Japan}
\begin{abstract}
The problem as to when two noncommuting observables are considered
to have the same value arises commonly, but shows a nontrivial difficulty.
Here, an answer is given by establishing the notion of perfect correlations
between noncommuting observables 
and applied to obtain a criterion for precise measurements of a given
observable in a given state.
\end{abstract}
\maketitle

% TEXT:

% 
  \newcommand{\beqa}{\begin{eqnarray}}
  \newcommand{\eeqa}{\end{eqnarray}}
  \newcommand{\beqas}{\begin{eqnarray*}}
  \newcommand{\eeqas}{\end{eqnarray*}}
% New Environments:
  \newtheorem{Theorem}{Theorem}
  \newenvironment{Proof}{\begin{trivlist}
    \item[\hskip \labelsep {\em \indent Proof.}]}{\qed\end{trivlist}}
 \newcommand{\qed}{{\em QED}}
  \newcommand*{\N}{\mathbf{N}}
  \newcommand*{\Q}{\mathbf{Q}}
  \newcommand*{\R}{\mathbf{R}}
  \newcommand*{\Z}{\mathbf{Z}}
  \newcommand*{\bA}{\mathbf{A}}
  \newcommand*{\bB}{\mathbf{B}}
  \newcommand*{\bC}{\mathbf{C}}
  \newcommand*{\bD}{\mathbf{D}}
  \newcommand*{\bE}{\mathbf{E}}
  \newcommand*{\bF}{\mathbf{F}}
  \newcommand*{\bG}{\mathbf{G}}
  \newcommand*{\bH}{\mathbf{H}}
  \newcommand*{\bI}{\mathbf{I}}
  \newcommand*{\bJ}{\mathbf{J}}
  \newcommand*{\bK}{\mathbf{K}}
  \newcommand*{\bL}{\mathbf{L}}
  \newcommand*{\bM}{\mathbf{M}}
  \newcommand*{\bN}{\mathbf{N}}
  \newcommand*{\bO}{\mathbf{O}}
  \newcommand*{\bP}{\mathbf{P}}
  \newcommand*{\bR}{\mathbf{R}}
  \newcommand*{\bS}{\mathbf{S}}
  \newcommand*{\bT}{\mathbf{T}}
  \newcommand*{\bV}{\mathbf{V}}
  \newcommand*{\bW}{\mathbf{W}}
  \newcommand*{\bX}{\mathbf{X}}
  \newcommand*{\bY}{\mathbf{Y}}
  \newcommand*{\bZ}{\mathbf{Z}}
  \newcommand*{\ba}{\mathbf{a}}
  \newcommand*{\bb}{\mathbf{b}}
  \newcommand*{\bc}{\mathbf{c}}
  \newcommand*{\bx}{\mathbf{x}}
  \newcommand*{\by}{\mathbf{y}}
  \newcommand*{\cA}{\mathcal{A}}
  \newcommand*{\cB}{\mathcal{B}}
  \newcommand*{\cC}{\mathcal{C}}
  \newcommand*{\cD}{\mathcal{D}}
  \newcommand*{\cE}{\mathcal{E}}
  \newcommand*{\cF}{\mathcal{F}}
  \newcommand*{\cG}{\mathcal{G}}
  \newcommand*{\cH}{\mathcal{H}}
  \newcommand*{\cI}{\mathcal{I}}
  \newcommand*{\cJ}{\mathcal{J}}
  \newcommand*{\cK}{\mathcal{K}}
  \newcommand*{\cL}{\mathcal{L}}
  \newcommand*{\cM}{\mathcal{M}}
  \newcommand*{\cN}{\mathcal{N}}
  \newcommand*{\cO}{\mathcal{O}}
  \newcommand*{\cP}{\mathcal{P}}
  \newcommand*{\cQ}{\mathcal{Q}}
  \newcommand*{\cR}{\mathcal{R}}
  \newcommand*{\cS}{\mathcal{S}}
  \newcommand*{\cT}{\mathcal{T}}
  \newcommand*{\cU}{\mathcal{U}}
  \newcommand*{\cV}{\mathcal{V}}
  \newcommand*{\cW}{\mathcal{W}}
  \newcommand*{\cX}{\mathcal{X}}
  \newcommand*{\cY}{\mathcal{Y}}
  \newcommand*{\cZ}{\mathcal{Z}}
  \newcommand*{\hA}{{\hat A}}
  \newcommand*{\hB}{{\hat B}}
  \newcommand*{\hC}{{\hat C}}
  \newcommand*{\hD}{{\hat D}}
  \newcommand*{\hE}{{\hat E}}
  \newcommand*{\hF}{{\hat F}}
  \newcommand*{\hG}{{\hat G}}
  \newcommand*{\hH}{{\hat H}}
  \newcommand*{\hI}{{\hat I}}
  \newcommand*{\hJ}{{\hat J}}
  \newcommand*{\hK}{{\hat K}}
  \newcommand*{\hL}{{\hat L}}
  \newcommand*{\hM}{{\hat M}}
  \newcommand*{\hN}{{\hat N}}
  \newcommand*{\hO}{{\hat O}}
  \newcommand*{\hP}{{\hat P}}
  \newcommand*{\hQ}{{\hat Q}}
  \newcommand*{\hR}{{\hat R}}
  \newcommand*{\hS}{{\hat S}}
  \newcommand*{\hT}{{\hat T}}
  \newcommand*{\hU}{{\hat U}}
  \newcommand*{\hV}{{\hat V}}
  \newcommand*{\hW}{{\hat W}}
  \newcommand*{\hX}{{\hat X}}
  \newcommand*{\hY}{{\hat Y}}
  \newcommand*{\hZ}{{\hat Z}}
   \newcommand*{\hn}{\hat{n}}
  \newcommand*{\hp}{\hat{p}}
 \newcommand*{\hq}{\hat{q}}
  \newcommand*{\hx}{\hat{x}}
  \newcommand*{\hy}{\hat{y}}
 \newcommand*{\oE}{\overline{E}}
  \newcommand*{\oQ}{\overline{Q}}
  \newcommand*{\oX}{\overline{X}}
  \newcommand*{\oY}{\overline{Y}}
  \newcommand*{\rA}{{\rm A}}
  \newcommand*{\rB}{{\rm B}}
  \newcommand*{\rC}{{\rm C}}
  \newcommand*{\rD}{{\rm D}}
  \newcommand*{\rE}{{\rm E}}
  \newcommand*{\rF}{{\rm F}}
  \newcommand*{\rG}{{\rm G}}
  \newcommand*{\rH}{{\rm H}}
  \newcommand*{\rI}{{\rm I}}
  \newcommand*{\rJ}{{\rm J}}
  \newcommand*{\rK}{{\rm K}}
  \newcommand*{\rL}{{\rm L}}
  \newcommand*{\rM}{{\rm M}}
  \newcommand*{\rN}{{\rm N}}
  \newcommand*{\rO}{{\rm O}}
  \newcommand*{\rP}{{\rm P}}
  \newcommand*{\rQ}{{\rm Q}}
  \newcommand*{\rR}{{\rm R}}
  \newcommand*{\rS}{{\rm S}}
  \newcommand*{\rT}{{\rm T}}
  \newcommand*{\rU}{{\rm U}}
  \newcommand*{\rV}{{\rm V}}
  \newcommand*{\rW}{{\rm W}}
  \newcommand*{\rX}{{\rm X}}
  \newcommand*{\rY}{{\rm Y}}
  \newcommand*{\rZ}{{\rm Z}}
  \newcommand*{\tA}{\tilde{A}}
  \newcommand*{\tB}{\tilde{B}}
  \newcommand*{\tC}{\tilde{C}}
  \newcommand*{\tD}{\tilde{D}}
  \newcommand*{\tE}{\tilde{E}}
  \newcommand*{\tF}{\tilde{F}}
  \newcommand*{\tG}{\tilde{G}}
  \newcommand*{\tH}{\tilde{H}}
  \newcommand*{\tI}{\tilde{I}}
  \newcommand*{\tJ}{\tilde{J}}
  \newcommand*{\tK}{\tilde{K}}
  \newcommand*{\tL}{\tilde{L}}
  \newcommand*{\tM}{\tilde{M}}
  \newcommand*{\tN}{\tilde{N}}
  \newcommand*{\tO}{\tilde{O}}
  \newcommand*{\tP}{\tilde{P}}
  \newcommand*{\tQ}{\tilde{Q}}
  \newcommand*{\tR}{\tilde{R}}
  \newcommand*{\tS}{\tilde{S}}
  \newcommand*{\tT}{\tilde{T}}
  \newcommand*{\tU}{\tilde{U}}
  \newcommand*{\tV}{\tilde{V}}
  \newcommand*{\tW}{\tilde{W}}
  \newcommand*{\tX}{\tilde{X}}
  \newcommand*{\tY}{\tilde{Y}}
  \newcommand*{\tZ}{\tilde{Z}}
  \newcommand*{\al}{\alpha}
  \newcommand*{\be}{\beta} 
  \newcommand*{\ch}{\chi}
  \newcommand*{\da}{\dagger}
  \newcommand*{\de}{\delta}
  \newcommand*{\ep}{\epsilon}
  \newcommand*{\et}{\eta}
  \newcommand*{\ga}{\gamma}
  \newcommand*{\io}{\iota}
  \newcommand*{\ka}{\kappa}
  \newcommand*{\la}{\lambda}
  \newcommand*{\mb}{\mbox}
  \newcommand*{\nn}{\nonumber}
  \newcommand*{\om}{\omega}
  \newcommand*{\ph}{\phi}
  \newcommand*{\ps}{\psi} 
  \newcommand*{\rh}{\rho}
  \newcommand*{\si}{\sigma} 
  \newcommand*{\ta}{\tau}
  \newcommand*{\up}{\upsilon}
  \newcommand*{\ve}{\varepsilon}
  \newcommand*{\vp}{\varphi}
  \newcommand*{\ze}{\zeta}
  \newcommand*{\De}{\Delta}                                          
  \newcommand*{\Ex}{{\rm Ex}}
  \newcommand*{\Eq}[1]{Eq.~(\ref{eq:#1})}
  \newcommand*{\Ga}{\Gamma}                                          
  \newcommand*{\IFF}{\Longleftrightarrow}
  \newcommand*{\IF}{\Longlefrarrow}
  \newcommand*{\Iff}{\Leftrightarrow}
  \newcommand*{\If}{\Leftarrow}
  \newcommand*{\La}{\Lambda}                                         
  \newcommand*{\Not}{\neg}
  \newcommand*{\Om}{\Omega}
  \newcommand*{\Or}{\vee}
  \newcommand*{\Ph}{\Phi}                                            
  \newcommand*{\Ps}{\Psi}                                            
  \newcommand*{\Si}{\Sigma}                                          
  \newcommand*{\THEN}{\Longrightarrow}
  \newcommand*{\Then}{\Rightarrow}
  \newcommand*{\Th}{\Theta}                                          
  \newcommand*{\Tr}{\mbox{\rm Tr}}
  \newcommand*{\Up}{\Upsilon}                                        
  \newcommand*{\Var}{{\rm Var}}
  \newcommand*{\eR}{\overline{\R}}
  \newcommand*{\eq}[1]{(\ref{eq:#1})}
\newcommand*{\bra}[1]{\langle#1|}
\newcommand*{\ket}[1]{|#1\rangle}
\newcommand*{\braket}[1]{\langle#1\rangle}
\newcommand*{\bracket}[1]{\langle#1\rangle}
\newcommand{\ketbra}[1]{\ket{#1}\bra{#1}}

\section{Introduction}

In quantum mechanics, we cannot predict a definite value of
a given observable generally, and it is sometimes stressed that
quantum mechanics does not speak of the value of an observable
in a single event, but only speaks of the average value over a
large number of events.
However, the quantum correlation definitely describes 
relations of values of observables in a single event as typically
in the EPR correlation,
where we cannot predict a definite value of 
the momentum or the position of each particle from an
EPR pair, whereas we can definitely predict the total momentum 
and the distance of the pair, and thereby we have a definite 
one-to-one correspondence between the values of their momenta
to be obtained from their joint measurements or between the values
of their positions.

In this paper, we shall investigate one of the most
fundamental aspects of quantum correlations; that is, 
we shall consider the general problem as to 
when two observables $X$ and $Y$ in a quantum system
can be considered to ``have the same value,'' in a given state,
 in the sense suggested above.  It should be stressed
that when we use this expression, we do not intend to make any
assumptions as to whether a definite value exists prior to the
measurement; such a question is a matter of the interpretation of
quantum mechanics and we do not enter into it.  Rather, we
choose to define what it means ``to have the same value'' in terms
of perfect correlations as the ones described above, meaning that,
if the two observables are jointly measured, one is guaranteed to
obtain the same value for both.  As we shall explain below, the
question of when two observables X and Y ``have the same value''
arises when one asks
if the time evolution changes the given observable 
and if an indirect measurement consisting of the measuring 
interaction and
the meter measurement is considered to precisely measure 
the given observable. 

For two classical random variables $X$ and $Y$, 
it is well accepted that $X$ and $Y$ have the same value 
if and only if $X$ and $Y$ are perfectly correlated, or
equivalently
the joint probability of obtaining different values of $X$ and $Y$
vanishes.
Thus, we can immediately generalize this notion to pairs of commuting
observables based on the well-defined joint probability distribution 
of commuting observables,
so that two commuting observables are considered to have the
same value in the given state if and only if they are perfectly correlated.
However, two operators are not necessarily commuting, 
and the generalization of the notion of perfect correlation 
to noncommuting
observables should be strongly demanded, whereas no serious
investigations  have been done. 
This paper introduces the notion of perfect correlations 
between arbitrary two observables, and characterizes it
by various statistical notions in quantum mechanics.
As a result, the above problems are shown to be answered
by simple and well-founded conditions
in the standard formalism of quantum mechanics.

\section{Difficulties in the notion of perfection correlation}
\label{se:2}

Let $A$ be an observable of a quantum system 
in a state $\ps$ at the origin of time.
Then, it is a fundamental question to ask whether 
the observable $A$ is unchanged or changed between 
two times $t_{1}$ and $t_{2}$.
Let $A(t)$ be the Heisenberg operator at time $t$
corresponding to the observable $A$.
If the question is asked independent of the system state $\ps$, 
the answer is that $A$ is unchanged if and only if $A(t_{1})=A(t_{2})$.
However, the question depending on the system state shows a 
nontrivial difficulty.

Let $D_{A}=A(t_{2})-A(t_{1})$ be the increment of $A$ from time
$t_{1}$ to $t_{2}$. Then, it is natural to expect that 
the value of the observable
$A$ is unchanged between two times $t_{1}$ and $t_{2}$ in
the system state $\ps$ if and only if the state $\ps$ is an
eigenstate of $D_{A}$ with eigenvalue 0, i.e., $D_{A}\ps=0$, or
equivalently
\beqa 
A(t_{1})\ps=A(t_{2})\ps.
\eeqa 
This means that the increment $D_{A}$ has the definite value
zero in the state $\ps$.  However, the above characterization is
unexpectedly not true in general. For example, let $A(t_{1})$
and $A(t_{2})$ be two $4\times 4$ matrices such that
\beqas
A(t_{1})=\left(
\begin{array}{cccc}
1 & 1 & 0 & 0\\
1 & 1& 0 & 0\\
0 & 0 & 1 & 1\\
0 & 0 & 1 & 0
\end{array}
\right),
\quad
A(t_{2})=\left(
\begin{array}{cccc}
1 & 1 & 0 & 0\\
1 &0& 0 & 0\\
0 & 0 & 1 & 1\\
0 & 0 & 1 &1
\end{array}
\right),
\eeqas
with time evolution operator 
$U(t_{2},t_{1})$ and the state $\ps$ such that
\beqas
U(t_{2},t_{1})=
\left(
\begin{array}{cccc}
0 & 0 & 1 & 0\\
0 & 0& 0 & 1\\
1 & 0 & 0 & 0\\
0 & 1 & 0 & 0
\end{array}
\right),
\quad
\ps=
\left(
\begin{array}{c}
 1 \\
0 \\
 0\\
 0
\end{array}
\right).
\eeqas
Then, we have $A(t_{1})\ps=A(t_{2})\ps$,
and hence the first and the second moments of $A$ are unchanged,
i.e., $\bracket{\ps|A(t_{1})|\ps}=\bracket{\ps|A(t_{2})|\ps}=1$
and $\bracket{\ps|A(t_{1})^{2}|\ps}=\bracket{\ps|A(t_{2})^{2}|\ps}
=2$.  However, we have
$\bracket{\ps|A(t_{1})^{3}|\ps}=4$ 
but $\bracket{\ps|A(t_{2})^{3}|\ps}=3$.
Thus, the third moment of $A$ is 
changed from time $t_{1}$ to $t_{2}$,
so that the observable $A$ is considered to have been
changed in this time interval.

On the other hand, the requirement that $A(t_{1})$ and $A(t_{2})$ should
have the same probability distribution in the state $\ps$ is a necessary but
not sufficient condition,
since there are cases where $A(t_{1})$ and $A(t_{2})$ have
the same probability distribution but they are statistically independent.
Specifically, suppose that $\ps$ is the product state of two copies of a
state $\ph$,   i.e., $\ps=\ph\otimes\ph$,  and that there is an
observable $B$ such that $A(t_{1})=B\otimes I$ and 
$A(t_{2})=I\otimes B$.
In this case, $A(t_{1})$ and $A(t_{2})$ have the same 
probability distribution,
but they are statistically independent in the case where
$\ph$ is not an eigenstate of $B$.  In fact,
$D_{A}\ps\not=0$ if and only if $\ph$ is not an eigenstate of $B$.  
Thus, in this case, we cannot judge
that the observable $A$ has been unchanged.

\section{Perfect correlation in measurement}

The notion of perfect correlation is not restricted  to the
problem on the Heisenberg time evolution, but also has broad
applications in foundations on quantum mechanics \cite{Per93}
and quantum information theory \cite{NC00}.  
Among them, another problem concerns
the notion of measurement. Any measurement has two
not necessarily commuting observables, one of which is the
observable to be measured and the other is the meter observable
after the measuring interaction  \cite{84QC,BLM91}.   
A fundamental question as to when
the given observable is precisely measured in a given state has
remained open.  However, this is obviously related to the perfect
correlation between the measured observable  and the meter
observable. This paper will solve this fundamental problem by
establishing the general notion of perfect correlations between
noncommuting observables.

Every measurement can be modeled by a process of 
indirect measurement described by the measuring interaction 
between the measured object  and the measuring 
apparatus followed by a subsequent observation of the meter 
observable in the apparatus
\cite{vN32,84QC,89RS,BLM91,BK92,NC00,04URN}.  
Let $\bS$ be
the object and $\bA$ the apparatus. Then, in order to measure the 
value of an observable $A$ in $\bS$ at time $t$,
the time of the measurement, the observer 
actually observes the value of the meter observable $M$ 
in $\bA$ at time $t+\De t$, where the measuring interaction is
supposed to be turned on from time $t$ to $t+\De t$.
Thus, in order to measure the observable $A(t)$,
the indirect measurement actually observes 
the observable $M(t+\De t)$.

A fundamental problem is to determine what condition
ensures that this measurement successfully measures 
the value of the observable $A$ at time $t$.
If we have a satisfactory notion of perfect correlation,
we can readily answer this question by stating that 
the indirect measurement successfully measures the
observable $A$ at time $t$ if and only if $A(t)$ and
$M(t+\De t)$ are perfectly correlated.
However, since $A(t)$ and $M(t+\De t)$ are not necessarily
commuting, the above question has not been answered
generally.

Instead, the conventional approach has questioned
what observable is measured by the above indirect measurement
independent of the input state.
Let $\ps$ be the state of $\bS$ at time $t$ and
$\xi$ the state of $\bA$ at time $t$.
We assume that the apparatus $\bA$ is always
prepared in the fixed state $\xi$ at the time 
of the measurement, while the object $\bS$ is 
in an arbitrary state $\ps$.
Then, the indirect measurement measures the 
observable $A$ at time $t$ if and only if
the two observables $A(t)$ and $M(t+\De t)$ have 
the same probability distribution for any state $\ps$
\cite{84QC,89RS,BLM91,BK92,NC00,04URN}.

Since the above definition of measurement 
of the observable independent of the input state 
does not explicitly require that $A(t)$ and $M(t+\De t)$ 
have the same value unless the measurement is carried out 
in an eigenstate of $A(t)$, it is not immediately obvious whether the value
randomly obtained by observing $M(t+\De t)$ would actually
correspond in any way to the value one would obtain, in the same situation, 
by an alternative (indirect or direct) measurement of $A(t)$.
Yet, there is something
unsatisfying about the possibility that, for any state not an eigenstate of
$A(t)$, the two operators $A(t)$ and $M(t+\De t)$ might just represent
independent random variables that just happen to have the same
distribution.  
One would certainly like to think that, in a precise
measurement, these two operators should "have the same value"---in the
sense defined in the Introduction---even under conditions when this value
may not be a definite quantity prior to the measurement. 

Indeed, the experimenter reads the value of the meter $M$ 
at time $t+\De t$ and records that the same value was taken 
by $A$ at time $t$; however, there might be a possibility that
another experimenter would obtained a different value of $A$
at time $t$ from another apparatus.
As above, it has not been ensured that this is not the case.

In order to solve the above problem,
this paper introduces the notion of perfect correlations 
between arbitrary two observables in any state, 
and characterizes it
by various statistical notions in quantum mechanics.
As a result, the above problem is affirmatively answered
by simple and well-founded conditions
in the standard formalism of quantum mechanics.
In particular, we shall establish a simple condition for 
the measured observable $A(t)$ and the meter observable
$M(t+\De t)$ to be perfectly correlated in a given state $\ps$,
and show that the conventional definition implies that the measured observable
$A(t)$ and the meter observable $M(t+\De t)$ are actually
perfectly correlated in any state $\ps$.
Thus, we shall conclude that 
the measured value from the meter observable
after the measuring interaction is not produced by
the interaction, but actually reproduces the value 
of the measured observable before the interaction.

\section{Definition of perfect correlations}
Let $X,Y$ be two observables in a quantum system $\bS$ described
by a Hilbert space $\cH$.
For simplicity, in this paper we assume that $\cH$ is finite dimensional.
The spectral projection $E^{X}(x)$ of $X$ for any $x\in\R$ is generally 
defined to be the projection operator of $\cH$ onto the subspace 
$\{\ps\in\cH|\ X\ps=x\ps\}$. 
If $X$ and $Y$ commute, their joint probability distribution in an
arbitrary state $\ps$ is defined by 
\beqa
\Pr\{X=x,Y=y\|\ps\}=\bracket{\ps|E^{X}(x)E^{Y}(y)|\ps}.
\eeqa
The above probability distribution is operationally 
interpreted as the joint probability distribution of the
measured values of $X$ and $Y$ in the
simultaneous measurement of $X$ and $Y$.
In general, we say that $X$ and $Y$ are {\em jointly distributed}
in state
$\ps$,   if 
\beqa\label{eq:joint-dist}
\bracket{\ps|E^{X}(x)E^{Y}(y)|\ps}\ge 0
\eeqa 
for any $x,y\in\R$. 
In this case, we have
\beqa\label{eq:030915h}
\bracket{\ps|E^{X}(x)E^{Y}(y)|\ps}=\bracket{\ps|E^{Y}(y)E^{X}(x)|\ps}.
\eeqa
Then, for any function $F(x,y)=\sum_{j,k}f_{j}(x)g_{k}(y)$ we have
\beqa
\lefteqn{
\sum_{x,y}F(x,y)\bracket{\ps|E^{X}(x)E^{Y}(y)|\ps}
}\quad\nn\\
&=&\bracket{\ps|\sum_{j,k}f_{j}(X)g_{k}(Y)|\ps}\\
&=&\bracket{\ps|\sum_{j,k}g_{k}(Y)f_{j}(X)|\ps}.
\eeqa

We say that $X$ and $Y$ are {\em perfectly correlated} in state $\ps$,
if 
\beqa\label{eq:def}
\bracket{\ps|E^{X}(x)E^{Y}(y)|\ps}=0
\eeqa
for any $x,y\in\R$ with $x\ne y$.
It is obvious that perfectly correlated observables are
jointly distributed.
Since $\bracket{\ps|E^{X}(x)|\ps}
=\sum_{y}\bracket{\ps|E^{X}(x)E^{Y}(y)|\ps}$,
the above condition is equivalent to the relation
\beqa\label{eq:PC-1}
\bracket{\ps|E^{X}(x)E^{Y}(y)|\ps}=\de_{x,y}\bracket{\ps|E^{X}(x)|\ps}
\eeqa
for any $x,y\in\R$,  where $\de_{x,y}$ stands for Kronecker's delta.
If $X$ and $Y$ are commuting, the above definition reduces to
the usual one that means that 
in the simultaneous measurement of $X$ and $Y$
the joint probability of the results 
$X=x$ and $Y=y$ vanishes, if $x\not=y$.
We shall show that a pair of observables $X,Y$ perfectly 
correlated in a state $\ps$ are considered to be 
simultaneously measurable in the state $\ps$
and that their outcomes always coincide each other.

We say that two observables $X$ and $Y$ are {\em equally
distributed}  in state $\ps$, if
$\bracket{\ps|E^{X}(x)|\ps}=\bracket{\ps|E^{Y}(x)|\ps}$ for all
$x\in\R$.
It follows easily from \Eq{PC-1} that  perfectly correlated
observables are equally distributed.
However, it is also obvious that the converse is not true 
even for commuting observables.

\section{Root mean square of difference}

Suppose that $X$ and $Y$ are perfectly correlated in $\ps$.
Then, intuitively speaking, they have the same value,
even though both of them are random.
Thus, it is expected that the difference $X-Y$ definitely has
the value zero, or equivalently $\ps$ is an eigenstate of 
$X-Y$ with eigenvalue 0, i.e, $X\ps=Y\ps$.  
In order to prove this property from our definition, 
we consider the distance $\|X\ps-Y\ps\|$ between
$X\ps$ and $Y\ps$. 
Obviously, $\|X\ps-Y\ps\|=0$ if and only if $X\ps=Y\ps$.
We generally have
\beqas
\lefteqn{\|X\ps-Y\ps\|^{2}}\\
&=&
\|\sum_{x}xE^{X}(x)\ps-\sum_{y}yE^{Y}(y)\ps\|^{2}\\
&=&
\sum_{x,y}(x-y)^{2}\Re\bracket{\ps|E^{X}(x)E^{Y}(y)|\ps}.
\eeqas
Thus, 
if  $X$ and $Y$ are jointly distributed in state $\ps$, we have
\beqa\label{eq:rms-distance}
\|X\ps-Y\ps\|^{2}=\sum_{x,y}(x-y)^{2}\bracket{\ps|E^{X}(x)E^{Y}(y)|\ps}.
\eeqa
Suppose that $X$ and $Y$ are perfectly correlated in state $\ps$.
Then, 
we have
\beqas
\lefteqn{\sum_{x,y}(x-y)^{2}\bracket{\ps|E^{X}(x)E^{Y}(y)|\ps}}\quad\\
&=&
\sum_{x,y}(x-y)^{2}\de_{x,y}\bracket{\ps|E^{X}(x)E^{Y}(y)|\ps}
=0,
\eeqas
so that \Eq{rms-distance} concludes $X\ps=Y\ps$.

Busch, Heinonen, and Lahti \cite{BHL04} showed that
the condition $X\ps=Y\ps$ does not imply that $X$ and
$Y$ are equally distributed.
Moreover, we have shown in Secion \ref{se:2} that this happens
even for unitarily equivalent observables $X$ and $Y$.
Thus, the condition $X\ps=Y\ps$ does not 
sufficiently characterize the
perfect correlation, even if $X$ and $Y$ have the same spectrum.
However, for jointly distributed 
$X$ and $Y$, the condition $X\ps=Y\ps$ implies their
perfect correlation.
To show this,
suppose that $X\ps=Y\ps$ and 
$X$ and $Y$ are jointly distributed in $\ps$.
Then, we have $\bracket{\ps|E^{X}(x)E^{Y}(y)|\ps}\ge 0$, 
and from \Eq{rms-distance} we have
$(x-y)^{2}\bracket{\ps|E^{X}(x)E^{Y}(y)|\ps}=0$ for any $x,y\in\R$.
Thus, we have $\bracket{\ps|E^{X}(x)E^{Y}(y)|\ps}=0$ if $x\not=y$,
and by definition $X$ and $Y$ are perfectly correlated in $\ps$.

Therefore, we have proven the following theorem.

\begin{Theorem}%\begin{Theorem}% 1.}
Two observables $X$ and $Y$ are perfectly correlated in state $\ps$
if and only if $X$ and $Y$ are jointly distributed and $X\ps=Y\ps$. 
\end{Theorem}

\section{Space of perfectly correlating states}

Suppose that $X$ and $Y$ are perfectly correlated in $\ps$.
It is natural to ask what states other than $\ps$ have this
property.
Since $X$ and $Y$ intuitively have the same value in $\ps$,
if we have obtained the result $X=x$ in measuring $X$
without disturbing $X$ and $Y$, 
we can also expect to have both $X=x$ and $Y=x$
in the state just after the above measurement.
Thus, it is natural to expect that $X$ and $Y$ are
perfectly correlated also in the state $E^{X}(x)\ps/\|E^{X}(x)\ps\|$
obtained by the above $X$ measurement,
and by linearity we can also expect that the state $f(X)\ps/\|f(X)\ps\|$
has this property.

In order to characterize all the states of the form $f(X)\ps/\|f(X)\ps\|$,
we introduce the following terminology.
The {\em cyclic subspace} spanned by an observable $X$ and a state $\ps$ 
is the subspace $\cC(X,\ps)$ spanned by $X^{n}\ps$ 
for any $n=0,1,2,\ldots$. 
It is easy to see that
$\cC(X,\ps)$ is the smallest $X$ invariant subspace of $\cH$ 
including $\ps$.  Denote by
$\cC_{1}(X,\ps)$ the unit sphere of
$\cC(X,\ps)$.
Denote by $P_{X,\ps}$ the projection of $\cH$ onto
$\cC(X,\ps)$.
Then, we have $f(X)P_{X,\ps}=P_{X,\ps}f(X)=P_{X,\ps}f(X)P_{X,\ps}$
for any function $f$.
Now, we have the following theorem.

\begin{Theorem}\label{th:2}
For any two observables $X$ and $Y$ and any state $\ps$,
the following conditions are equivalent.

(i) Observables $X$ and $Y$ are perfectly correlated in state $\ps$.

(ii) Observables $X$ and $Y$ are perfectly correlated in any state 
$\ph\in\cC_{1}(X,\ps)$.

(iii) $f(X)\ps=f(Y)\ps$ for any function $f$.

(iv) $f(X)P_{X,\ps}=f(Y)P_{X,\ps}$

(v) $XP_{X,\ps}=YP_{X,\ps}$.
\end{Theorem}

\begin{Proof}
Suppose that condition (i) holds.  
By the similar computations as before, we have
$\|f(X)\ps-f(Y)\ps\|^{2}=0$, and hence, 
the implication (i)$\Then$(iii) follows. 
Suppose that condition (iii) holds. 
Then, we have
$f(X)g(X)\ps=f(Y)g(Y)\ps=f(Y)g(X)\ps$ for any $f$ and $g$. Since every
$\ph\in\cC(X,\ps)$ is of the form $\ph=g(X)\ps$ for some $g$, we have
$f(X)P_{X,\ps}=g(Y)P_{X,\ps}$. Thus, the implication (iii)$\Then$(iv)
follows. The implication (iv)$\Then$(v) is obvious.
Suppose that condition (v) holds.
Let $P=P_{X,\ps}$.
Since  $X$ leaves $\cC(X,\ps)$ invariant, so does $Y$. 
Thus, the spectral projections of $YP$ and $XP$
on $\cC(X,\ps)$ are $E^{Y}(y)P$ and $E^{X}(y)P$, 
respectively, and hence
$E^{Y}(y)P=E^{X}(y)P$ for any $y\in\R$,
so that $E^{X}(x)E^{Y}(y)P=E^{X}(x)E^{X}(y)P$.
Thus, we have
$\bracket{\ph|E^{X}(x)E^{Y}(y)|\ph}=0$, if $x\not=y$, for any
$\ph\in\cC_{1}(X,\ps)$. 
It follows that $X$ and $Y$ are perfectly correlated
in any state $\ph\in\cC_{1}(X,\ps)$.   
Thus, the implication (v) $\Then$ (ii) has been proven.
Since the implication (ii) $\Then$ (i) is obvious, the proof is completed. 
\end{Proof}

By the above theorem, observables $X$ and $Y$ are represented 
on the space $\cC(X,\ps)$ by the same operator $XP_{X,\ps}=
YP_{X,\ps}$, and hence $X$ and $Y$ are considered
to be simultaneously measurable in $\ps$ and to have
the identical outcomes.
In fact, if one measures $X$ and $Y$ by consecutive projective 
measurements of $X$ and $Y$, then by Theorem \ref{th:2} (iv) the joint
probability distribution of the two outcomes satisfies
$$
\|E^{Y}(y)E^{X}(x)\ps\|^{2}=
\|E^{X}(y)E^{X}(x)\ps\|^{2}=
\de_{x,y}\bracket{\ps|E^{X}(x)|\ps},
$$
and hence the measurement outputs actually show
the perfect correlation predicted by
the theoretical joint probability distribution \eq{PC-1}.

\section{Characterization of perfectly correlating states}
From the above theorem we have the following important
characterization of perfectly correlating states.

\begin{Theorem}% 3.}
Two observables $X$ and $Y$ are perfectly correlated
in a state $\ps$ if and only if $\ps$ is a superposition
of common eigenstates of $X$ and $Y$ with common
eigenvalues.
\end{Theorem}

\begin{Proof}
Suppose that $X$ and $Y$ are perfectly correlated in a
state $\ps$.  Then, $\cC(X,\ps)$
is generated by eigenstates of $XP_{X,\ps}=YP_{X,\ps}$.
Thus, $\ps$ is a superposition of common eigenstates of 
$X$ and $Y$ with common eigenvalues.
Conversely, suppose that  $\ps$ is a superposition
of common eigenstates of $X$ and $Y$ with common
eigenvalues.
Then, the subspace $\cS$ generated by those eigenstates is
invariant under both $X$ and $Y$ and includes $\ps$.
Thus, $\cC(X,\ps)\subset\cS$, and $X=Y$ on $\cC(X,\ps)$,
and hence from Theorem 2 (v), we conclude $X$  and $Y$
are perfectly correlated in $\ps$.
\end{Proof}

\section{Equally distributed observables}
Theorem 2 (ii) suggests that perfectly correlated 
$X$ and $Y$ in $\ps$ are equally 
distributed in any state in the cyclic subspace spanned 
by $\ps$ and $X$.
The following theorem shows that the converse is also true.

\begin{Theorem}% 4.}
Two observables $X$ and $Y$ are perfectly correlated in state $\ps$
if and only if they are equally distributed in any state $\ph$ in
$\cC_{1}(X,\ps)$.
\end{Theorem}

\begin{Proof}
Suppose that $X$ and $Y$ are perfectly correlated in state $\ps$.
From Theorem 2 (iv), we have $f(X)\ph=f(Y)\ph$
for any function $f$ and $\ph\in\cC(X,\ps)$.
Taking $f$ to be $f(y)=\de_{x,y}$, we have
$
\bracket{\ph|E^{X}(x)|\ph}
=\bracket{\ph|E^{Y}(x)|\ph},
$
so that $X$ and $Y$ are equally distributed for any $\ph\in\cC_{1}(X,\ps)$.
Conversely, suppose that $X$ and $Y$ are equally distributed 
in any state $\ph$ in $\cC_{1}(X,\ps)$.
There is an orthonormal
basis $\{\ket{n,\nu}\}$ of $\cC(X,\ps)$ consisting of eigenstates
of $X$ such that $X\ket{n,\nu}=x_{n}\ket{n,\nu}$.
By the equal distributivity of $X$ and $Y$ in $\ket{n,\nu}$, 
we have $Y\ket{n,\nu}=x_{n}\ket{n,\nu}$. 
Thus, $\ps$ is a superposition of common eigenstates of
$X$ and $Y$ with common eigenvalues.
We conclude, therefore, from Theorem 3 that
$X$ and $Y$ are perfectly correlated in state $\ps$.
\end{Proof}

\section{Characterization of precise measurements of observables}

Let $\bA(\bx)$ be an apparatus with output variable
$\bx$ for measuring a system $\bS$ described
by a Hilbert space $\cH$.
The measuring process of $\bA(\bx)$ is described by
a quadruple $(\cK,\xi,U,M)$ consisting of a 
Hilbert space $\cK$ describing the probe $\bP$,
a state vector $\xi$ in $\cK$ describing the state of $\bP$
just before the measurement,
a unitary operator $U$ on $\cH\otimes\cK$ describing the
time evolution of the composite system $\bS+\bP$
during the measuring interaction,
and an observable $M$ on $\cK$ describing the meter observable
\cite{84QC,89RS,BLM91,BK92,00MN,01OD,04URN}.
We assume for simplicity that both $\cH$ and $\cK$ are
finite dimensional.
If the measuring interaction turns on from time $t$ to $t+\De t$,
in the Heisenberg picture with original state $\ps\otimes\xi$
at time $t$, we write $A(t)=A\otimes I$ and 
$M(t+\De t)=U^{\da}(I\otimes M)U$.

The probability distribution of the output variable $\bx$
on the input state $\ps$ is given by
\beqa
\Pr\{\bx=x\| \ps\}
=\bracket{\ps\otimes\xi|U^{\da}[I\otimes E^{M}(x)]U|\ps\otimes\xi}.
\eeqa
Let $A$ be an observable on $\cH$.
Naturally, we should say that the apparatus $\bA(\bx)$ 
with measuring process
$(\cK,\xi,U,M)$ {\em precisely measures} 
the value of observable $A$ in state $\ps$,
if the observable $A\otimes I$ and $U^{\da}(I\otimes M)U$ 
are perfectly correlated in the state $\ps\otimes\xi$. 
In this case, we can say that the measuring interaction reproduces
``the value'' taken by $A$ before the measuring interaction; 
if the observer were to measure $A(t)$ and $M(t+\De t)$ jointly 
then the observer would obtain the same value from each measurement,
so that the observer can safely report that his value obtained from
observing $M(t+\De t)$ is the value obtained from the measurement
of $A(t)$.
On the other hand, the apparatus $\bA(\bx)$ is said to satisfy 
the {\em Born statistical formula (BSF)} for $A$ in state $\ps$ if
\beqa
\Pr\{\bx=x\|\ \ps\}=\bracket{\ps|E^{A}(x)|\ps}
\eeqa
for all $x\in\R$.
In this case, we can say at least that the measuring interaction 
reproduces the probability distribution of $A$ before the measuring
interaction.

The relation
\beqa
\Pi(x)=\Tr_{\cK}[U^{\da}[I\otimes
E^{M}(x)]U(I\otimes\ket{\xi}\bra{\xi})]
\eeqa
defines the  {\em probability operator valued measure
(POVM)} $\{\Pi(x)|\ x\in\R\}$ of $\bA(\bx)$,
where $\Tr_{\cK}$ stands for the partial trace over $\cK$.
Then, the probability distribution of the output is described
by
\beqa
\Pr\{\bx=x\|\ \ps\}=\bracket{\ps|\Pi(x)|\ps}.
\eeqa
We say that a POVM $\{\Pi(x)|\ x\in\R\}$ is {\em perfectly
correlated} to an observable $A$ in a state $\ps$, if
\beqa\label{eq:PC-POVM}
\bracket{\ps|\Pi(x)E^{A}(y)|\ps}=0
\eeqa
for any $x,y\in\R$ with $x\not=y$.
Then, the following theorem characterizes 
precise measurements of the value of an observable in a given state.

\begin{Theorem}% 5.}
Let $\bA(\bx)$ be an apparatus
with measuring process $(\cK,\xi,U,M)$ 
and POVM $\{\Pi(x)|\ x\in\R\}$.
Then, for any observable $A$ and state $\ps$,
the following conditions are all equivalent.

(i) $\bA(\bx)$ precisely measures $A$ in $\ps$.

(ii) The POVM $\{\Pi(x)|\ x\in\R\}$ is perfectly
correlated to $A$ in $\ps$.

(iii) $\bA(\bx)$ satisfies the BSF for $A$ in any $\ph\in\cC_{1}(A,\ps)$.

(iv) $\Pi(x)P_{A,\ps}=E^{A}(x)P_{A,\ps}$ for any $x\in\R$.
\end{Theorem}

\begin{Proof}
The equivalence between conditions (i) and (ii) follows immediately
from the relation
\beqa
\lefteqn{
\bracket{\ps\otimes\xi|[E^{A}(x)\otimes I]U^{\da}[I\otimes E^{M}(y)]U|
\ps\otimes\xi}}\hspace{10em}\nn\\
&=&
\bracket{\ps|E^{A}(x)\Pi(y)|\ps}.\quad
\eeqa
We easily obtain the relations
\beqa
&\cC(A\otimes I,\ps\otimes\xi)=\cC(A,\ps)\otimes\bC\xi,&
\label{eq:030915a}\\
&P_{A\otimes I,\ps\otimes\xi}=P_{A,\ps}\otimes\ket{\xi}\bra{\xi}.&
\label{eq:030915b}
\eeqa
From the above relations, the equivalence of conditions (i) and (iii)
follows from Theorem 4. Assume that condition (i) holds.
By Theorem 2, condition (i) is equivalent to the relation
\beqa\label{eq:030915d}
U^{\da}[I\otimes E^{M}(x)]UP_{A\otimes I,\ps\otimes\xi}
=[E^{A}(x)\otimes I]P_{A\otimes I,\ps\otimes\xi}
\eeqa
for any $x\in\R$.
Then, $U^{\da}[I\otimes E^{M}(x)]U$ commutes with 
$P_{A\otimes I,\ps\otimes\xi}$, so that from \Eq{030915b}
we have 
\beqa
U^{\da}[I\otimes E^{M}(x)]UP_{A\otimes I,\ps\otimes\xi}
=
\Pi(x)P_{A,\ps}\otimes\ket{\xi}\bra{\xi}.
\eeqa
Thus, \Eq{030915d} implies the relation
\beqa
\Pi(x)P_{A,\ps}\otimes\ket{\xi}\bra{\xi}
=E^{A}(x)P_{A,\ps}\otimes\ket{\xi}\bra{\xi},
\eeqa
so that we have condition (iv).
Conversely, it is now easy to see that condition (iv)
implies \Eq{030915d}.
Thus, condition (i) and condition (iv) are equivalent.
\end{Proof}

The above theorem shows that whether an apparatus 
precisely measures the value of an observable in a given state is 
determined solely by the corresponding POVM. 
In the conventional approach, the apparatus
$\bA(\bx)$ is said to {\em precisely measure} the ``observable''
$A$, if it satisfies the BSF for $A$ in {\em every} state $\ps$ of the
system $\bS$ \cite{84QC,89RS,BLM91,BK92,NC00,04URN}.
It is well-known that $\bA(\bx)$ precisely
measures $A$ if and only if $\Pi(x)=E^{A}(x)$ for all $x\in\R$.
By Theorem 5, 
$\bA(\bx)$ satisfies the BSF for $A$ in
{\em every} state $\ps$ of the measured system if and only if the meter
observable and the measured observable are perfectly correlated in any
input state. Thus, we have justified the conventional 
definition by having shown that every precise measurement of 
``observable'' $A$ reproduces not only the probability distribution 
but also the value taken by $A$ before the measurement.

\section{von Neumann's  model of measurement}

It was shown by von Neumann \cite{vN55} 
that a measurement of an observable 
\beqa\label{eq:A}
A=\sum_{n}a_{n}\ketbra{\ph_{n}}
\eeqa
 on $\cH$
with eigenvalues $a_{0},a_{1},\ldots$ and an orthnormal basis
of eigenvectors
$\ph_{0},\ph_{1},\ldots$ can be realized by a unitary operator 
$U$ on the tensor product $\cH\otimes\cK$  with another separable
Hilbert space $\cK$ with orthonormal basis $\{\xi_{n}\}$ 
such that
\beqa\label{eq:U}
U(\ph_{n}\otimes\xi)=\ph_{n}\otimes\xi_{n},
\eeqa
where $\xi$ is an arbitrary vector state in $\cK$.
Let 
\beqa\label{eq:B}
M=\sum_{n}a_{n}\ketbra{\xi_{n}}
\eeqa
be an observable on $\cK$ called the meter.
von Neumann's model defines an apparatus $\bA(\bx)$ with
measuring process $(\cK,\xi,U,M)$.

Let us suppose that the initial state of the system is given by an arbitrary
state vector $\ps=\sum_{n}c_{n}\ph_{n}$.
Then, it follows from the linearity of $U$ we have
\beqa\label{eq:U-sp}
U(\ps\otimes\xi)=\sum_{n}c_{n}\ph_{n}\otimes\xi_{n}.
\eeqa
The conventional explanation as to why this transformation can be
regarded as a measurement is as follows; symbols are adapted to
the present context in the quote below.  ``In the state \eq{U-sp},
obtained by the measurement, there is a statistical correlation between
the state of the object and that of the apparatus: the simultaneous 
measurement on the system---object-plus-apparatus---of the two quantities,
one of which is the originally measured quantity of the object and the
second the position of the pointer of the apparatus, always leads to concordant
results.  As a result, one of these measurements is unnecessary:  The state
of the object can be ascertained by an observation on the apparatus.
This is a consequence of the special form of the state vector \eq{U-sp},
on not containing any $\ph_{m}\otimes\xi_{n}$ term with $n\not=m$
\cite{Wig63}.''
``The equations of motion permit the description of the process whereby
the state of the object is mirrored by the state of an apparatus.
The problem of a measurement on the object is thereby transformed into
the problem of an observation on the apparatus \cite{Wig63}.''

The above explanation correctly points out the existence of the 
statistical correlation between the measured observable $A$ and
the meter observable $M$ in the state \eq{U-sp}.  However,
this is not the statistical correlation between the measured observable
before the interaction and the meter observable after the interaction,
but that between those observables after the interaction.
Thus, the above statistical correlation does not even ensure that
the probability distribution of the measured observable before the
interaction is reproduced by the observation of the meter observable
after the interaction.

The role of the measuring interaction described by
$U$ should be to make the following two correlations:
(i) the correlation between the measured observable
$A$ {\em before} the interaction and the meter 
$M$ {\em after} the interaction, and (ii) the correlation between
the meter $M$ {\em after} the interaction and 
the measured observable $A$ {\em after} the interaction.
The first correlation is required by the {\em value reproducing requirement}
that the interaction transfers the value of the measured observable
$A$ before the interaction to the value of the meter $M$ after
the interaction.
The second correlation is required by the {\em repeatability hypothesis}
that if the meter observable $M$ has the value $a_{n}$ after the
interaction, then the observable $A$ also have the same value $a_{n}$
after the interaction so that the second measurement of $A$ after the
interaction reproduce the same value of the meter of the first
measurement od $A$.  

Now, we shall show that those requirements are actually satisfied.
Let $\et_{0},\et_{1},\ldots$ be an orthonormal basis of $\cH$ such that 
$\et_{0}=\xi$, namely an orthonormal basis extending $\{\xi\}$.
Let $\Ps_{n,m}$ be a unit vector in $\cH$ defined by
$\Ps_{n,m}=U^{\da}(\ph_{n}\otimes\xi_{m})$ for any $n,m$.
Then, we have $\Ps_{n,n}=\ph_{n}\otimes\xi$ and the 
family $\{\Ps_{n,m}\}$ is an orthonormal basis of $\cH$.
By simple calculations, we have
\beqa
A\otimes I
&=&A\otimes\ketbra{\xi}
+\sum_{m\not=0}A\otimes\ketbra{\et_{m}},\\ 
U^{\da}(A\otimes I)U
&=&A\otimes\ketbra{\xi}
+\sum_{n\not=m}a_{n}\ketbra{\Ps_{n,m}},\\
U^{\da}(I\otimes M)U
&=&A\otimes\ketbra{\xi}
+\sum_{n\not=m}a_{m}\ketbra{\Ps_{n,m}},
\eeqa
where $\sum_{n\not=m}$ stands for the summation over all
$n,m$ with $n\not=m$.  By the above relations it is now obvious that
$A\otimes I=U^{\da}(A\otimes I)U=U^{\da}(I\otimes M)U$
on their common invariant subspace $\cH\otimes [\xi]$,
so that those three observables are perfectly correlated in
the state $\ps\otimes\xi$ for every state vector $\ps$ in $\cH$.
Therefore, von Neumann's model $(\cK,\xi,U,M)$ satisfies both
the the value reproducing requirement and the repeatability
hypothesis.

\section{Concluding remarks}

In this paper, we have introduced the notion of perfect correlation 
between
noncommuting observables and explored its basic properties. 
This notion is applied to characterizing the precise measurement
of the value of an observable in a given state and justifies
the conventional definition of precise measurement of an
observable formulated independently of the input state.
Although this paper has focussed on the finite level systems,
the theory for the general case can be developed with analogous
results under the definition that observables $X$ and $Y$ are 
{\em perfectly correlated} in state $\ps$, if 
\begin{equation}
\bracket{\ps|E^{X}(\De)E^{Y}(\Ga)|\ps}=0
\end{equation}
for any mutually disjoint Borel sets $\De$ and $\Ga$, 
where $E^{X}$ and
$E^{Y}$ are the spectral measures of $X$ and $Y$, 
respectively; the detail
will be discussed in a forthcoming paper.

The notion of perfect correlation is not restricted to the
problem of measurement, but also has broad
applications in foundations on quantum mechanics \cite{Per93}
and quantum information theory \cite{NC00}.  
Those applications will be discussed elsewhere.

\section*{Acknowledgements}
The author thanks Julio Gea-Banacloche for his helpful
suggestions for revising the original manuscript.
This work was supported by the SCOPE project
of the MPHPT of Japan,  by the CREST
project of the JST, and by the Grant-in-Aid for Scientific Research of
the JSPS.
%\bibliographystyle{elsart-num}
%\bibliography{myelist,myebib}

\end{document}